\documentclass[10pt, conference]{IEEEtran}

\usepackage{tikz}
\usepackage{amsmath}

\usepackage{filecontents}
\usepackage{tikz}
\usepackage{filecontents}
\usepackage{cite}
\usepackage{amsmath,amsfonts,mathrsfs,amssymb}
\usepackage{graphicx,subfig}
\usepackage[normalem]{ulem}
\usepackage[linesnumbered,vlined,boxed, commentsnumbered, ruled]{algorithm2e}
\usepackage[ruled,linesnumbered]{algorithm2e}
\usepackage{multirow}
\usepackage{amsopn}
\usepackage{etoolbox}
\usepackage{xcolor}
\usepackage{url}

\usepackage{xspace}
\usepackage{indentfirst}
\usepackage{breakurl}
\usepackage{microtype}
\usepackage{color}
\usepackage{listings}
\usepackage{enumitem}
\usepackage{diagbox}
\usepackage{colortbl}
\usepackage{pifont}
\usepackage{amsthm}
\usepackage{makecell}

\def\BibTeX{{\rm B\kern-.05em{\sc i\kern-.025em b}\kern-.08em
    T\kern-.1667em\lower.7ex\hbox{E}\kern-.125emX}}

\newcommand{\etc}{\emph{etc.}\xspace}
\newcommand{\ie}{\emph{i.e.,}\xspace}
\newcommand{\eg}{\emph{e.g.,}\xspace}

\newcommand{\todo}[1]{\textbf{\textcolor{red}{(TODO:)#1}}}
\newcommand{\name}{{\sc Phoenix}\xspace}

\begin{document}

\title{\huge In-Orbit Processing or Not? Sunlight-Aware Task Scheduling for Energy-Efficient Space Edge Computing Networks}
\author{
	\IEEEauthorblockN{Weisen Liu$^{\dag}$, Zeqi Lai$^{\dag\ddag}$, Qian Wu$^{\dag\ddag}$, Hewu Li$^{\dag\ddag}$, Qi Zhang$^{\ddag}$, Zonglun Li$^{\dag}$, Yuanjie Li$^{\dag\ddag}$, Jun Liu$^{\dag\ddag}$}
	\IEEEauthorblockA{
		\textit{$^{\dag}$Institute for Network Sciences and Cyberspace, BNRist, Tsinghua University, Beijing 100084, China} \\
		\textit{$^{\ddag}$Zhongguancun Laboratory, Beijing, China}
	}
} 

\maketitle
\begin{abstract}

With the rapid evolution of space-borne  capabilities, \emph{space edge computing~(SEC)} is becoming a new computation paradigm for future integrated space and terrestrial networks. Satellite edges adopt advanced on-board hardware, which not only enables new opportunities to perform complex intelligent tasks in orbit, but also involves new challenges due to the additional energy consumption in power-constrained space environment.

In this paper, we present \name, an energy-efficient task scheduling framework for emerging SEC networks. \name exploits a key insight that in the SEC network, there always exist a number of \emph{sunlit edges} which are illuminated during the entire orbital period and have sufficient energy supplement from the sun. \name accomplishes energy-efficient in-orbit computing by judiciously offloading space tasks to ``sunlight-sufficient'' edges or to the ground. Specifically, \name first formulates the \emph{SEC battery energy optimizing~(SBEO)} problem which aims at minimizing the average battery energy consumption while satisfying various task completion constraints. Then \name incorporates a sunlight-aware scheduling mechanism to solve the SBEO problem and schedule SEC tasks efficiently. Finally, we implement a \name prototype and build an SEC testbed. Extensive data-driven evaluations demonstrate that as compared to other state-of-the-art solutions, \name can effectively reduce up to 54.8\% SEC battery energy consumption and prolong battery lifetime to $2.9\times$ while still completing tasks on time.

\end{abstract}
\section{Introduction}
\label{sec:introduction}

With the rapid evolution in the aerospace industry, emerging low earth orbit~(LEO) satellite mega-constellations not only extend the network boundary of today's terrestrial Internet, but also spawn an innovative computation paradigm: \textbf{``space edge computing~(SEC)''}. Based on advanced on-board hardware, emerging SEC technologies combine the capabilities of satellite communication and edge computing to provide edge-like services right at the satellite~\cite{yang2022algorithm,buttar2022semantic,cheng2023an}, and further enable a series of intelligent space applications such as smart remote sensing~\cite{intelligent_remote_sensing}, autonomous debris detection and avoidance~\cite{app12105106}, and Internet of space things~\cite{osuwa2017application}, \etc

While SEC has broad application prospects, it also involves new technical challenges in the energy-constrained outer space environment. On the one hand, fully realizing the promising capability of SEC requires extra advanced on-board hardware to support complex space missions, \eg exploiting satellite GPUs~\cite{denby2020orbital,denby2023kodan} for data inference and deploying high-speed inter-satellite communication links~(ISL) for cross-edge collaborative processing~\cite{zhang2023energy}. On the other hand, these additional payloads inevitably involve more energy consumption on satellite edges.
Since the energy usage can significantly affect the execution of SEC tasks as well as the battery lifetime~(as we will introduce later in \S\ref{sec:background}), \emph{accomplishing energy-efficient space task execution is undoubtedly a crucial problem for futuristic SEC networks.}

Task offloading, which has been well-studied by the conventional mobile computing community over the past decade, is an effective approach for optimizing energy consumption on power-constrained devices~\cite{guo2016energy,kwak2015dream}. The core idea behind traditional task offloading is to transfer the energy-intensive tasks from the power-constrained devices to a high-performance server with sufficient power supplement~(\eg a cloud), and receive the results after remote execution. Energy can be saved if the network transmission consumes less energy than local task execution. In an SEC scenario, a straightforward method to apply task offloading for energy-efficiency is to transfer space tasks to a nearby ground station which typically has sufficient computation capability and power supplement. However, due to huge amount of SEC data~\cite{vasisht2021l2d2}, naively offloading all tasks to the ground can easily overwhelm the satellite downlink~\cite{denby2023kodan} and involve high latency which is unacceptable for time-sensitive SEC tasks like satellite-based wildfire monitoring and rescue.

To address the limitation of existing offloading approaches, this paper presents \name, an energy-efficient and deadline-aware task scheduling framework for emerging SEC networks. The design of \name stands upon a series of important insights obtained from today's LEO satellite constellations. First, sub-systems in a satellite edge are powered directly by solar panels with sufficient power supplement when the satellite is illuminated by sunlight, and are powered by a rechargeable battery when the satellite enters the earth's shadow. Thus, the key to energy optimization is to reduce the energy consumption of the satellite battery. Second, as satellites move, we observe that there dynamically exist a number of orbit planes where satellites in these orbits are exposed to sunlight with near-100\% sunlit ratio. Carrying out computation tasks on these ``sunlit edges'' does not consume battery power. Third, an SEC task is typically associated with a time-to-completion requirement, especially for time-sensitive applications. Taken them together, \name accomplishes energy-efficiency by dynamically and judiciously offloading space tasks to computational nodes with sufficient power supplement subjecting to various task deadlines. To this end, \name makes scheduling decisions based on the following options: (i) processing the data locally on a satellite edge; (ii) offloading SEC tasks to other proper ``sunlit edges''; or (iii) offloading SEC tasks to ground stations. 


Specifically, \name calculates the decisions in two steps. First, given the SEC network information and the completion time requirements of various tasks, \name formulates the \emph{SEC Battery Energy Optimization~(SBEO)} problem which targets at minimizing the maximum depth-of-discharge~(DoD) of all satellite edge batteries, while satisfying various network, computation and application-level constraints. DoD is an important metric that characterizes the energy usage of a battery and can affect the lifetime of the rechargeable battery as well as the satellite itself. However, efficiently solving the SBEO problem is non-trivial since we prove its NP-hardness and multiple concurrent space tasks can compete for the dynamic computation and network resources in the SEC network.

Second, \name incorporates a sunlight-aware dynamic SEC task scheduling mechanism which decomposes the original SBEO problem and adopts a series of heuristic algorithms to calculate appropriate scheduling decisions efficiently. Specifically, to reduce the problem complexity, \name jointly combines coarse-grained orbit-level and find-grained per-satellite task scheduling to calculate near-optimal task assignments.

We build a data-driven hardware-in-the-loop SEC testbed and implement a \name prototype upon it. Our testbed integrates large-scale SEC network simulation, and low-power computational hardware that has been verified in real space environments.  Extensive evaluations demonstrate that \name can outperform other state-of-the-art SEC approaches in terms of energy consumption, battery lifetime, task deadline satisfaction  \etc, under various experiment configurations.

Contributions of this paper can be summarized as follows: (i) we formulate the SEC battery energy optimization~(SBEO) problem and expose the technical challenges of solving it efficiently and effectively; (ii) we propose \name, a novel sunlight-aware energy-efficient task scheduling framework for optimizing satellite battery usage and extending the lifetime of SEC networks; (iii) we implement a \name prototype and conduct extensive data-driven, hardware-in-the-loop experiments to demonstrate the effectiveness of \name.

\section{Technical Background}
\label{sec:background}

\subsection{Space Edge: A New Intelligent Computing Paradigm}
\label{subsec:space_edge_computing}

\noindent
\textbf{On-board hardware evolution in aerospace industry.} 
On-board network and computation capabilities have increased significantly over the past decade. The high-speed inter-satellite and ground-satellite links~(\ie ISLs and GSLs) have been successfully deployed on satellite platforms, which can offer Gbps-level data transmission for satellite communication~\cite{gregory2012commercial,carrizo2020optical} and facilitate the inter-networking of a large number of satellites.
Besides, on-board processing capability has also been promoted via low-power commercial off-the-shelf~(COTS) hardware accelerators, \eg graphics processing unit~(GPU)~\cite{denby2023kodan}, vision processing unit~(VPU)~\cite{giuffrida2021varphi}, and field programmable gate array~(FPGA)~\cite{rapuano2021fpga}. In 2020, European Space Agency~(ESA) launched the $\Phi$-sat-1 sensing satellite equipped with Intel Movidius Myriad 2 to verify the on-board performance of deep neural network~(DNN) and achieved great success~\cite{buckley2022radiation}.

\noindent
\textbf{Space edge computing~(SEC) enables in-orbit intelligence.} With such space hardware evolution above, recently we have witnessed a new intelligent computation paradigm: \emph{space edge computing~(SEC)}. SEC exploits edge-like computing capabilities on satellites and within satellite networks to process data in orbits. SEC can be used in many innovative space applications such as autonomous remote decisions, in-orbit detection for disaster discovery, climate monitoring and maritime rescue~\cite{giuffrida2020cloudscout,wei2020hrsid,spiller2022wildfire}. SEC enables faster data processing, reduced latency, and improved efficiency by handling data in orbit rather than sending all data back to the ground.

\begin{figure}[tbp]
	\centering
	\includegraphics[width=0.99\linewidth]{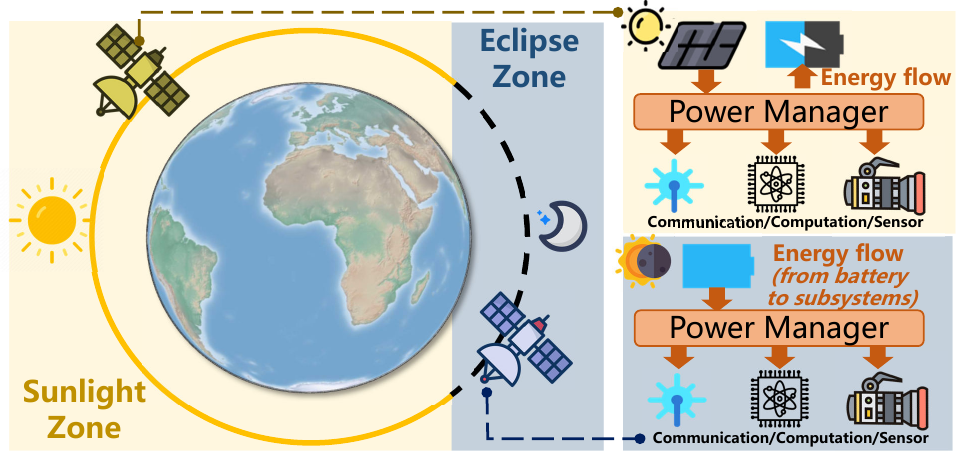}
	\vspace{-0.2in}
	\caption{A typical architecture of SEC on-board power system.}
	\vspace{-0.25in}
	\label{fig:power_model}
\end{figure}

\begin{figure}[t]
	\centering
	\subfloat[Impact of DoD on cycle number.]{
		\includegraphics[width=0.47\linewidth]{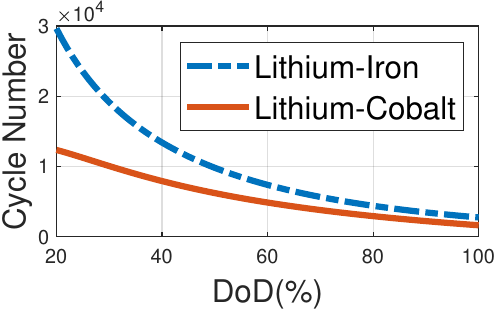}
		\label{fig:instant_local_processing}
	}
	\subfloat[Impact of DoD on battery lifetime.]{
		\includegraphics[width=0.47\linewidth]{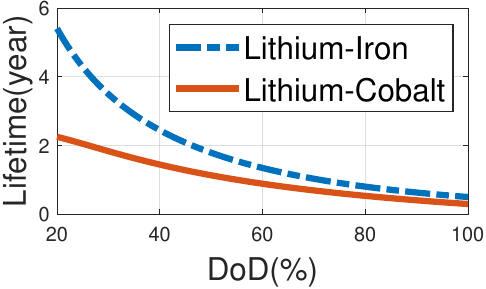}
		\label{fig:delayed_local_processing}
	}
	\vspace{-0.05in}
	\caption{The impact of DoD on the total number of charging-discharging cycles and battery lifetime.}
	\vspace{-0.25in}
	\label{fig:lifetime_vs_dod}
\end{figure}

\subsection{Battery Energy Consumption: The Achilles' Heel to SEC}
\label{subsec:energy_consumption_of_sec}

However, while the advanced on-board hardware provides intelligence, it also involves additional energy consumption, which not only affects the execution of SEC tasks, but also affects the lifetime of the SEC system itself.

\noindent
\textbf{Satellite power supplement.}
Satellites orbit around the earth, and during a portion of their orbit, they enter the earth's shadow, causing an eclipse.
Fig.~\ref{fig:power_model} plots a typical architecture of existing SEC power systems~(SPS).
During the sunlight phase, solar panels generate energy from the received photons, and SPS distributes power to other subsystems of the satellite. On-board rechargeable batteries store excess power generated by solar panels when the satellite is in sunlight, and the stored energy is used to power the satellite during eclipse periods.
Typically, among all subsystems the communication and computation modules can contribute copious energy consumption~\cite{denby2020orbital}.

\noindent
\textbf{Depth of discharge~(DoD) and battery lifetime.} The lifetime of a battery~(as well as the satellite itself) is tightly affected by its power usage during the charging-discharging cycles of the battery. In particular, DoD characterizes the percentage of discharged energy relative to the maximum capacity. 
DoD is equal to 0 if the battery is full and 100\% if the battery is empty. 
As the battery is used, the maximum capacity of the battery will gradually decay. Typically, when the maximum battery capacity falls below 80\% of its initial volume, the battery should be retired, which is defined as its lifetime~\cite{yang2020state,hu2020battery}. 
Previous works~\cite{mallon2017analysis,li2015analysis} have uncovered that if we regularly discharge a battery at a higher DoD, its lifetime will be shorter (\eg about 20\% DoD increase can reduce the lifetime by half). 
Fig.~\ref{fig:lifetime_vs_dod} shows how DoD affects the total number of available charging-discharging cycles and the lifetime of two kinds of Lithium batteries. 
Considering that battery usage can jointly affect task performance and satellite lifespan, optimizing the battery energy consumption and accomplishing energy-efficient in-orbit task processing is important for futuristic SEC networks.

\section{\name Design Overview}
\label{sec:overview}

In this section, we introduce the key idea and overview of our sunlight-aware energy-efficient task scheduling strategy.

\subsection{Observation: Sunlight-Sufficient Space Edges}
\label{subsec:observation_sunlit_sufficient_space_edge}

Based on the SPS knowledge introduced in \S\ref{subsec:energy_consumption_of_sec}, we know that the key to optimizing energy usage and prolonging lifetime for SECs is to reduce the \emph{battery energy consumption} caused by in-orbit task processing. To this end, our \name design leverages an important observation: in an LEO satellite constellation, typically there exist many \emph{``sunlight-sufficient''} satellites which spend most of their orbital time in sunlight.

\begin{figure}[t]
	\centering
	\subfloat[CDF of sunlit ratio.]{
		\includegraphics[width=0.5\linewidth]{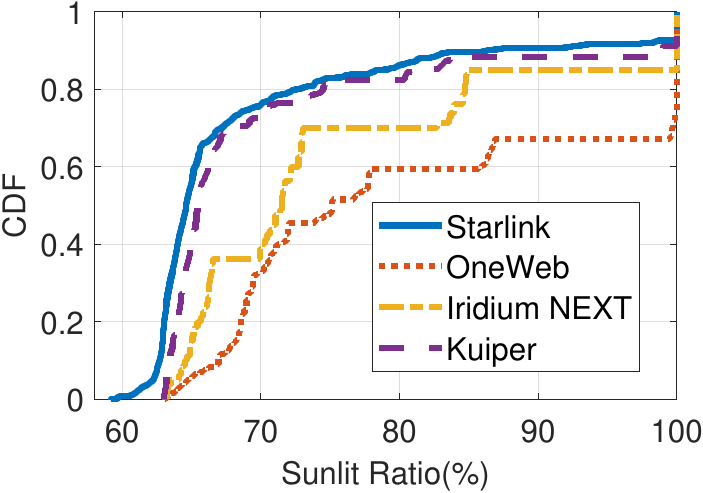}
		\label{fig:sunlit_ratio_cdf}
	}
	\subfloat[Example of 100\% sunlit ratio.]{
		\includegraphics[width=0.4\linewidth]{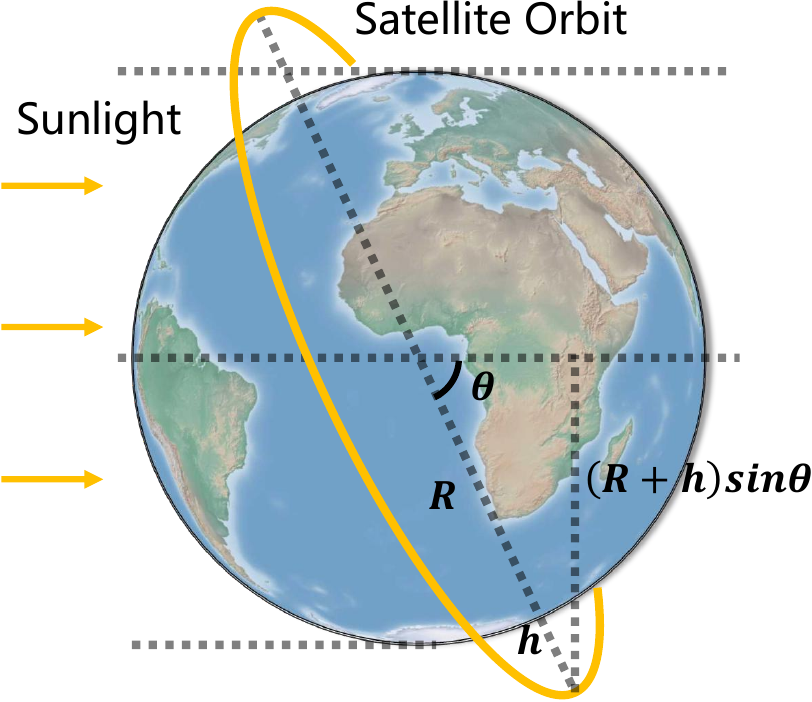}
		\label{fig:sunlit_whole_period_example}
	}
	\vspace{-0.05in}
	\caption{Sunlit ratio analysis for representative constellations.}
	\vspace{-0.25in}
\end{figure}

To quantitatively introduce and understand this phenomenon, we define a metric called \emph{sunlit ratio}, which is calculated by the ratio of the period a satellite is in sunlight to its total operation period. Fig.~\ref{fig:sunlit_ratio_cdf} plots the CDF of sunlit ratio of four state-of-the-art LEO satellite constellations which differ in their orbital altitudes and inclinations. We calculate the sunlit ratio based on their public real-world satellite trajectories~\cite{tle_data} during May 2023. We obtain three important observations. First, we find that the sunlit ratio is approximately higher than 60\%, indicating that a large fraction of satellites are exposed to the sun for a long time on their orbits. Second, interestingly we observe that some satellites in certain orbits can achieve near-100\% sunlit ratio with sufficient sunlit supplement. Fig.\ref{fig:sunlit_whole_period_example} plots an example to explain this observation. Suppose that radius of the earth is $R$. The height of satellite orbit is denoted as $h$ and the angle between sunlight and orbital plane is denoted as $\theta$. Then, the distance between the orbit and the earth's core is $R+h$ and the vertical component of distance perpendicular to sunlight is $(R+h)\cdot sin\theta$. If the length of the vertical component is longer than the radius of the earth, the whole orbit can be exposed to the sunlight and the sunlit ratio can reach 100\%.

\subsection{Basic Idea: Sunlight-Aware SEC Task Scheduling}
\label{subsec:basic_idea_sunlit_task_scheduling}

\begin{figure}[t]
	\centering
	\subfloat[Real-time local processing.]{
		\includegraphics[width=0.45\linewidth]{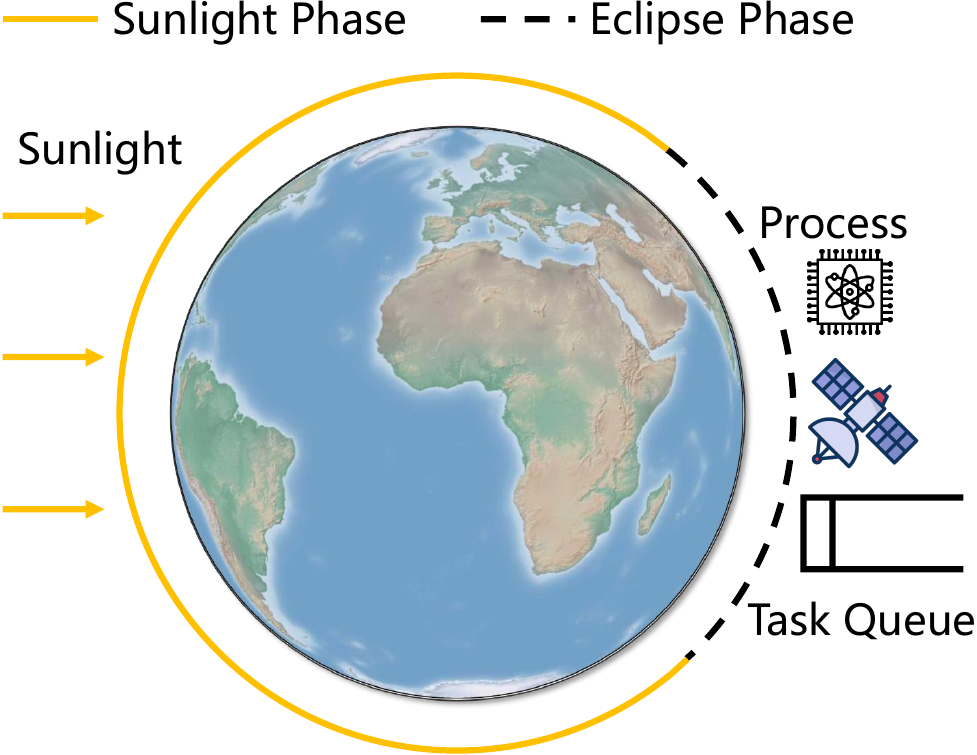}
		\label{fig:instant_local_processing}
	}
	\subfloat[Delayed local processing.]{
		\includegraphics[width=0.45\linewidth]{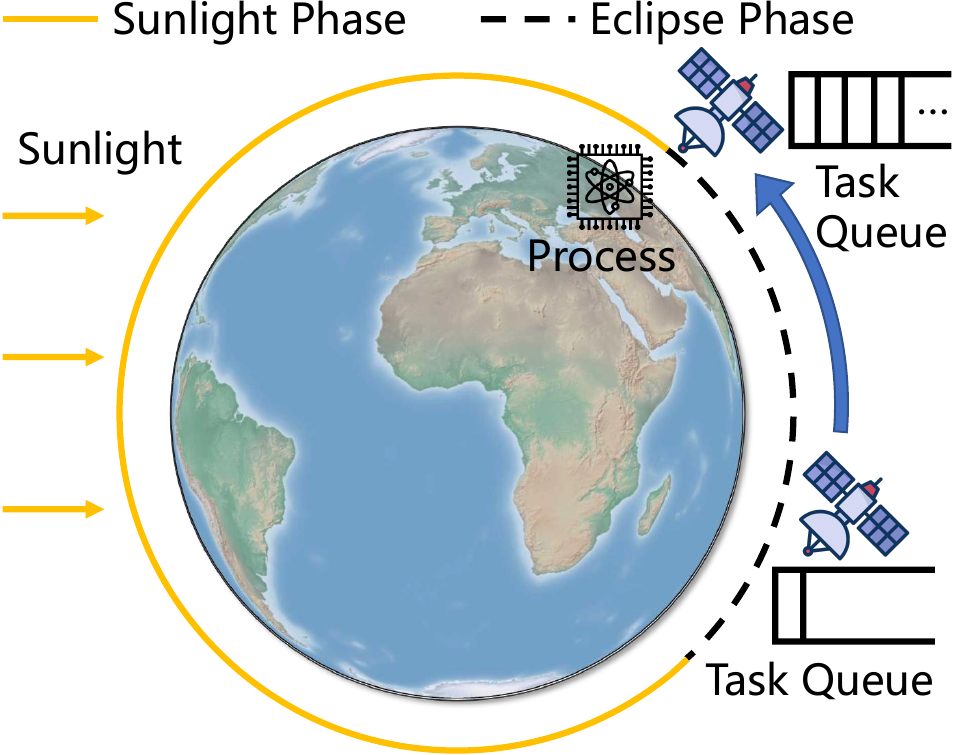}
		\label{fig:delayed_local_processing}
	}
	\vspace{-0.1in}
	\subfloat[Satellite offloading.]{
		\includegraphics[width=0.45\linewidth]{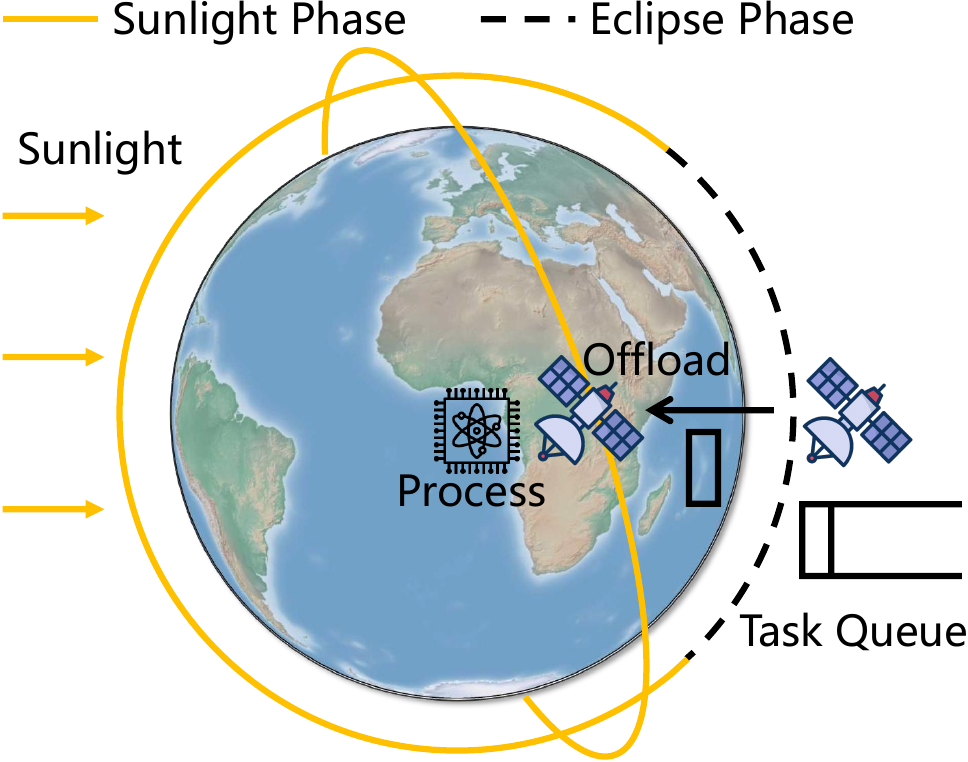}
		\label{fig:instant_satellite_offloading}
	}
	\subfloat[Ground station offloading.]{
		\includegraphics[width=0.45\linewidth]{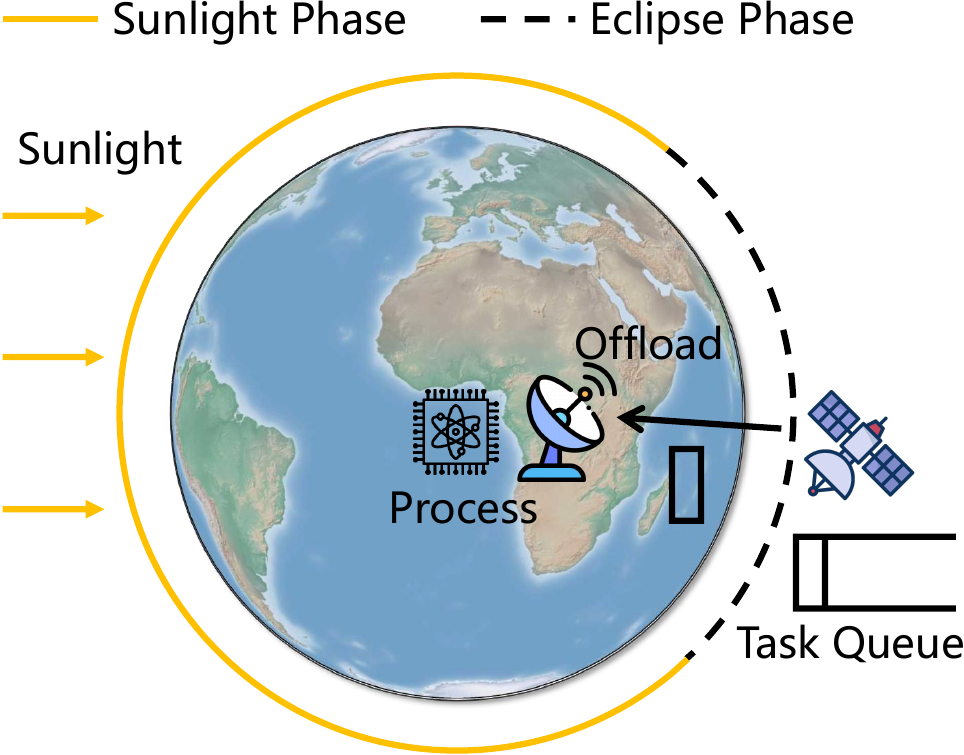}
		\label{fig:ground_station_offloading}
	}
	\vspace{-0.05in}
	\caption{The trade-space of SEC task scheduling in \name.}
	\label{fig:offloading_strategies_explanation}
	\vspace{-0.2in}
\end{figure}

Inspired by the above crucial observations, \name exploits a key idea to accomplish energy-efficient SEC: \emph{dynamically offloading in-orbit tasks to appropriate power-sufficient nodes~(\eg sunlight-sufficient satellites with near-100\% sunlit ratio) without exceeding various task deadlines.} Specifically, \name judiciously schedules SEC tasks based on the following options~(also illustrated in Fig.~\ref{fig:offloading_strategies_explanation}) to minimize the battery energy consumption while meeting various requirements of task completion time:
\begin{itemize}[leftmargin=*]
	\item \textbf{Real-time local processing}~(Fig.~\ref{fig:instant_local_processing}). Once in-orbit data is acquired on a satellite, processing it locally and immediately no matter where the current satellite is.
	\item \textbf{Delayed local processing}~(Fig.~\ref{fig:delayed_local_processing}). Once in-orbit data is acquired, the satellite delays the data processing until a specific time point~(\eg when the satellite leaves eclipse).
	\item \textbf{Offloading tasks to nearby sunlit edges}~(Fig.~\ref{fig:instant_satellite_offloading}). Instead of being processed locally, space data collected in orbit is transferred to another sunlight-sufficient satellite in the SEC network for energy-efficient processing.
	\item \textbf{Offloading tasks to available ground stations}~(Fig.~\ref{fig:ground_station_offloading}). Space data is transferred to a ground station with sufficient power and computation capability through the SEC network.
\end{itemize}

Essentially, the above scheduling options represent different preferences on saving battery energy consumption and guaranteeing task completion time. For example, immediate local data processing may achieve shorter mission completion time, but possibly at the cost of higher battery energy consumption. Transferring the entire in-orbit task to the ground may save energy for a satellite edge, but offloading high-volume space data can easily overwhelm the limited downlink and involve unacceptable latency. \name dynamically schedules various SEC tasks under different computation, network and sunlight conditions, and further makes proper scheduling decisions.

\subsection{System Overview}
\label{subsec:system_overview}

\begin{figure}[t]
	\centering
	\includegraphics[width=0.99\linewidth]{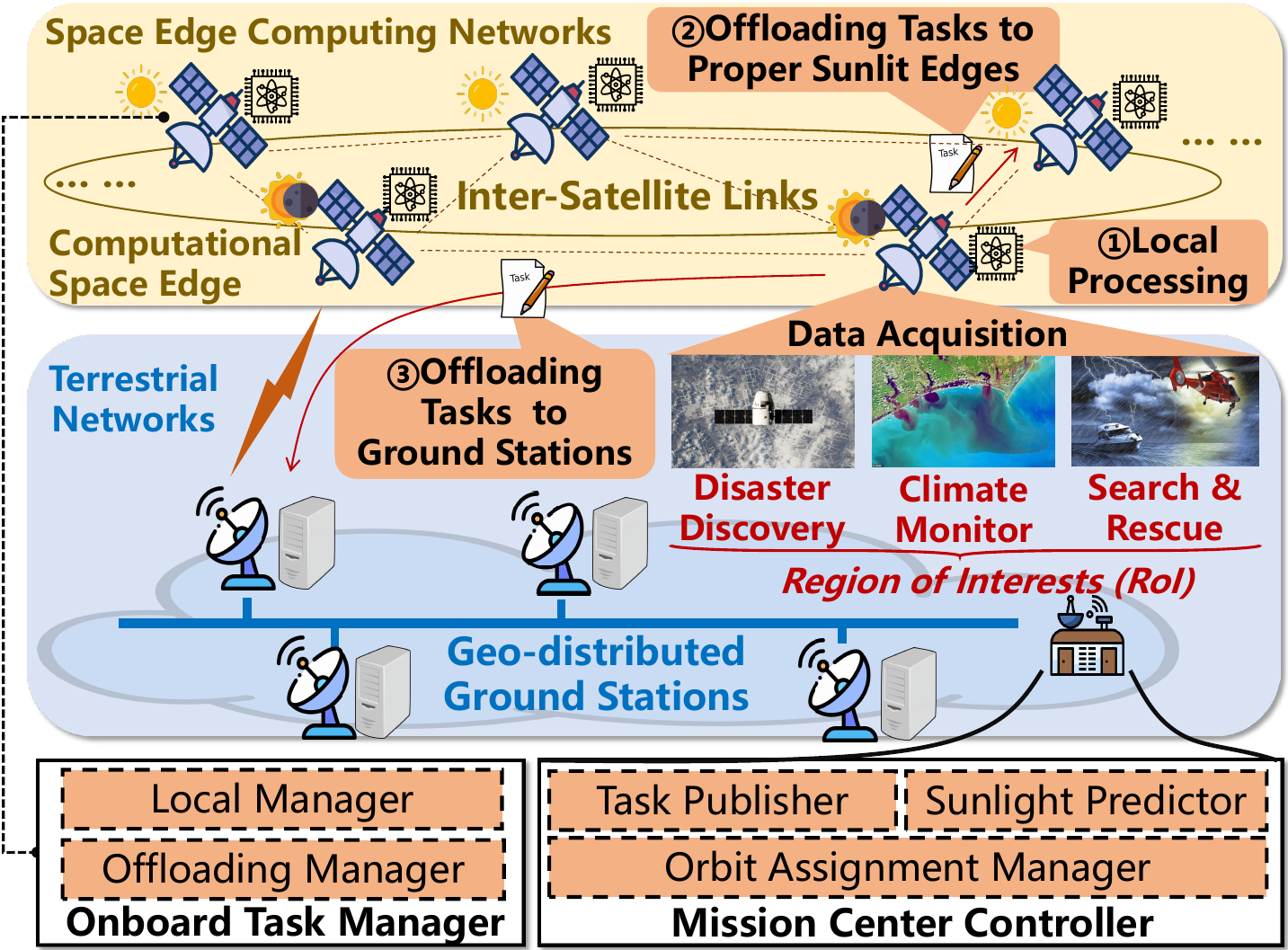}
	\vspace{-0.15in}
	\caption{\name system overview.}
	\vspace{-0.25in}
	\label{fig:system_architecture}
\end{figure}

\noindent
\textbf{Architecture.} 
Fig.~\ref{fig:system_architecture} shows the system overview of \name,
which combines: (i) a swarm of computational space edges constructing an SEC network, and (ii) terrestrial infrastructures such as geo-distributed ground stations and a mission control center. Each satellite is equipped with high-speed ISLs and GSLs for inter-satellite and ground-satellite communication. In addition, an on-board intelligent processor is deployed on each satellite for in-orbit computing. Based on this baseline SEC architecture, \name accomplishes energy-efficient SEC task scheduling by incorporating two new components as follows.
\begin{itemize}[leftmargin=*]
	\item \textbf{Centralized mission center controller.} The controller is a centralized coordinator, which consists of three modules: task publisher, sunlight predictor and orbit assignment manager. The task publisher distributes the tasks to satellites. The sunlight predictor predicts the sunlight states of satellites based on their trajectories and distributes the sunlight information to satellites for offloading decisions. The orbit assignment manager pre-allocates alternative orbit sets for satellites to avoid resource competition between orbits.
	\item \textbf{Distributed on-board task manager.} The manager on each satellite consists of two modules: local processing manager and offloading manager. The offloading manager decides where to process the task based on the pre-allocated orbit set. The local processing manager arranges the task execution time if a task is decided to be processed locally.
\end{itemize}


\noindent
\textbf{Workflow.} SEC tasks are scheduled by the following steps: 
\begin{itemize}[leftmargin=*]
	\item \textbf{Task publication and orbit assignment.} The mission center receives tasks from customers, such as requirements of using SEC for disaster discovery, climate monitoring, maritime search and rescue. The common feature of these tasks is that they all require a constellation of satellites persistently capture information~(\eg high-resolution images) of a certain region of interest~(RoI). Then, the mission center predicts the sunlight states of satellites and publishes the RoI along with sunlight information to all satellites. Based on the sunlight and task information, the mission center pre-assigns an orbit subset for each orbit and distributes the assignment decision.
	\item \textbf{Task offloading.} When a satellite flys over RoI, it captures images and selects offloading destination for each task, either a ground station or a satellite (including itself, \ie local processing). The satellite first checks whether the task can be offloaded to a ground station before deadline. 
	If not, it tries to arrange the task locally and judges whether it can finish the task under sunlight phase as well as before deadline. If both of the above conditions can not be satisfied, it offloads the task to other satellite.
	\item \textbf{Processing time arrangement.} The final destination satellite arranges the task processing time (the ground station can process the task immediately after offloading) and sends the result back to the mission center when the task is finished. 
\end{itemize}

\section{Energy-Efficient Task Scheduling by \name}
\label{sec:design}

In this section, we introduce \name's sunlight-aware energy-efficient task scheduling mechanism in detail. 


\subsection{Formulating The SEC Battery Energy Optimization Problem}
\label{subsec:problem_formulation}

\noindent
\textbf{SEC network topology.} 
Assume that time is slotted with each timeslot $\Delta T$ and the time set is denoted as $\mathcal{T}=\{1,2,\dots,T_{max}\}$. Denote $G_t=(V,E_t)$ as the satellite network topology. $V$ is the node set and $E_t$ is the edge set at timeslot $t$. There are two types of nodes, \ie satellites and ground stations. Assume that the total number of orbits is $M$ and the satellite set in orbit $i$ can be denoted as $ORB_{i}$. Thus, the satellite of the whole constellation can be represented by $SAT=\cup_{1\leq i\leq M}ORB_{i}$. The number of ground stations is $N$ and the ground station set is denoted as $GS=\{g_1,g_2,\dots,g_N\}$. Then the node set can be described by $V=SAT\cup GS$. 
As satellites orbit around the earth, the visibility between nodes changes over time. Let binary variable $Vis_{i,j}^{t}$ denote the visibility between node $i$ and node $j$. If $i$ and $j$ are visible at time $t$, then $Vis_{i,j}^{t}=1$ and they can establish a link $(i,j)$. Therefore, we can check whether a satellite $s$ can connect to any ground station at time $t$ by a binary value $l_{s,t}$:
\begin{align}
	l_{s,t} = 1-\prod_{g\in GS}(1-Vis_{s,g}^{t}).		
\end{align}
Denote $\lambda$ as the number of ISLs in each satellite. Typically, there are four ISLs and one GSL for each satellite~\cite{bhattacherjee2019network}.
The link capacity of $(i,j)$ is denoted as $Cap(i,j)$ and we set $Cap(i,i)=\infty$, meaning that if a task is processed locally, 
the transmission can be finished in situ instantly.  

\noindent
\textbf{Task scheduling.} Assume that the task set is $TASK$ and a task $k$ can be represented by a tuple $(src_{k},z_{k},T_{arv}^{k},T_{cp}^{k},T_{ddl}^{k})$, indicating the source satellite creating the task, the task size, the task arrival time, the computing time required to process the task for satellite, and the deadline respectively. 
A task can be offloaded to ground stations, processed locally or offloaded to another satellite. Let $dst_k$ denote the node where task $k$ is processed. If $dst_k\in GS$, task $k$ is downloaded by ground stations directly; if $dst_k=src_k$, task $k$ is processed locally; otherwise, task $k$ is offloaded to another satellite. 
Then, the tasks that select node $s$ as the processing destination can be represented by $\mathcal{K}_{s}=\{k|dst_{k}=s,k\in TASK\}$.
Let $T_{of}^{k}$ be the point of time when task $k$ finishes offloading.
The task offloading time is decided by the bottleneck of path from $src_{k}$ to $dst_{k}$. Given a routing path $R(i,j,t)$ from node $i$ to node $j$ at timeslot $t$, 
the offloading path can be denoted as $R(src_{k},dst_{k},t)$. We use a binary variable $r_{i,j}^{k,t}$ to indicate whether the offloading flow of $k$ goes through link $(i,j)$ at $t$:
\begin{align}
	r_{i,j}^{k,t} = 
	\begin{cases}
		1, (i,j)\in R(src_{k},dst_{k},t)\\
		0, otherwise
	\end{cases}.
\end{align}
Assume that all flows on the same link share the link capacity fairly. Then, the bandwidth that each flow can obtain on link $(i,j)$ at timeslot $t$ is $Cap(i,j)/\sum_{k\in TASK}r_{i,j}^{k,t}$. 
Denote $sz_{k,t}$ as the data size of task $k$ transmitted at timeslot $t$. The data size that can be transmitted in each timeslot is limited by the minimum bandwidth on the offloading path:
\begin{align}
	sz_{k,t} = \Delta T \cdot \min\{\frac{Cap(i,j)}{\sum_{\sigma \in TASK}r_{i,j}^{\sigma,t}}|r_{i,j}^{k,t}=1, i,j\in V\}.
\end{align}
When the amount of sent data reaches the task size, the offloading is finished and the time to finish offloading is:
\begin{align}
	T_{of}^{k} = \min\{\tau|\sum_{t=T_{arv}^{k}}^{\tau} sz_{k,t}\geq z_{k}, \tau\in \mathcal{T}\}.
\end{align}
After the task is offloaded, the receiver should decide when to process the task. 
Let $T_{bcp}^{k}$ denote the time to begin processing and the time to complete processing can be represented by $T_{bcp}^{k}+T_{cp}^{k}-1$.
We use a binary variable $x_{k,t}$ to indicate whether the task $k$ is being processed at time $t$:
\begin{align}
	x_{k,t}=
	\begin{cases}
		1, & T_{bcp}^{k}\leq t\leq T_{bcp}^{k}+T_{cp}^{k}-1\\
		0, & otherwise
	\end{cases}.
\end{align}
Thus, we denote the task processing matrix as $\mathcal{X}=\{x_{k,t}| k\in TASK,t\in \mathcal{T}\}$. 
For any satellite $s$, the number of tasks under processing at timeslot $t$ can be calculated by $\sum_{k\in \mathcal{K}_{s}}x_{k,t}$.

\noindent
\textbf{Energy Consumption.} As mentioned in \S~\ref{subsec:energy_consumption_of_sec}, the electronic components in satellites consist of solar panels, communication terminals, processing units, battery and other basic modules (\eg sensors). For satellite $s$, we denote the power generated by its solar panels under sunlight phase as $P_{solar}^{s}$. As satellites enter eclipse phase and sunlight phase alternately, we use binary variable $sun_{s,t}$ to indicate whether satellite $s$ is illuminated at time $t$. Thus, the power generated by solar panels at timeslot $t$ is $sun_{s,t}\cdot P_{solar}^{s}$. 
We use $P_{ISL}^{s}$ and $P_{GSL}^{s}$ to describe the transmit power of ISL and GSL respectively. Therefore, the power consumed by ISL and GSL at timeslot $t$ are $\lambda\cdot P_{ISL}^{s}$ and $l_{s,t}\cdot P_{GSL}^{s}$ respectively.
Denote the power of processing units as $P_{cp}^{s}$. 
Thus, the power consumed by processing units at timeslot $t$ is $P_{cp}^{s}\cdot \sum_{k\in \mathcal{K}}x_{k,t}$. The power consumed by other basic modules is denoted as $P_{basic}^{s}$. 
Let $B_{vol}^{s}$ be the volume of battery and $B_{s,t}$ be the rest battery energy of $s$ at time $t$, which is decided by $\mathcal{K}_{s}$ and $\mathcal{X}$. In the beginning, the battery is full, \ie $B_{s,0}=B_{vol}^{s}$. As time goes by, the rest battery energy changes according to the energy produced by solar panels and the energy consumed by electronic devices:
\begin{align}
	B_{s,t}=&\min\{(sun_{s,t}\cdot P_{solar}^{s}-P_{basic}^{s}-P_{cp}^{s}\cdot \sum_{k\in \mathcal{K}_{s}}x_{k,t} \notag\\
	&-\lambda\cdot P_{ISL}^{s}-l_{s,t}\cdot P_{GSL}^{s})\cdot \Delta T + B_{s,t-1},B_{vol}^{s}\}.
\end{align}
Therefore, the DoD of satellite $s$ at timeslot $t$ can be represented by $1-B_{s,t}/B_{vol}^{s}$.

\noindent
\textbf{SEC Battery Energy Optimization~(SBEO) problem formulation.} Given the following inputs: (i) time set $\mathcal{T}$ and timeslot duration $\Delta T$; (ii) node set $V$ and visibility $Vis_{i,j}^{t}$ between nodes; (iii) link capacity $Cap(i,j)$; (iv) task set $TASK$ and routes between nodes $R(i,j,t)$; (v) power of electronic devices $P_{solar}^{s}$, $P_{basic}^{s}$, $P_{ISL}^{s}$, $P_{GSL}^{s}$ and $P_{cp}^{s}$; (vi) battery volume $B_{vol}^{s}$ and sunlight indicator $sun_{s,t}$, we aim to provide the processing task set $\mathcal{K}_{s}$ for all satellites and the processing matrix $\mathcal{X}$ such that the maximum DoD among all satellites is minimized, prolonging the lifetime of satellites as illustrated in \S~\ref{subsec:energy_consumption_of_sec}:
\begin{align}
	\min \max_{s\in SAT,t\in \mathcal{T}} 1-B_{s,t}/B_{vol}^{s}
\end{align}

\noindent
Constraints:

\noindent
(i) A satellite can only process a task at each timeslot:
\begin{align}
	\sum_{k\in \mathcal{K}_{s}}x_{k,t}\leq 1,\forall s\in SAT,t\in\mathcal{T}.
\end{align}

\noindent
(ii) Task processing should be scheduled after offloading finish:
\begin{align}
	T_{of}^{k}\leq T_{bcp}^{k}, \forall k\in TASK.
\end{align}

\noindent
(iii) Task processing should be completed before deadline:
\begin{align}
	T_{bcp}^{k}+T_{cp}^{k}\leq T_{ddl}^{k},\forall k\in TASK.
\end{align}

\noindent
\textbf{Complexity analysis.} To analyze the complexity, we simplify the SBEO problem to an easier case where ground stations are not considered and there is no transmission cost. Then, a task $k$ can be divided into $T_{cp}^{k}$ slots and a satellite $s$ can be divided into $T_{max}$ slots. There are two types of satellite slots according to the sunlight state, \ie sunlight slots and eclipse slots. If a task slot is assigned to a eclipse slot of a satellite, the cost is 1 while there is no cost when assigned to a sunlight slot. Then, the problem can be transformed into a generalized assignment problem, which assigns the task slots to satellite slots, aiming to minimize the cost. The generalized assignment problem has been proven to be NP-hard~\cite{nauss2003solving}. Thus, the simplistic problem as well as the SBEO problem are NP-hard.

\setlength{\textfloatsep}{0.12cm}
\begin{algorithm}[tb]
    \caption{Sunlight-aware Orbit Assignment} 
    \label{alg:orbit_assignment}
	\DontPrintSemicolon
	\KwIn{topology $G_t$, task set $TASK$, indicator $sun_{s,t}$}
	\KwOut{alternative subset $alt\_set$}
	$cyc\leftarrow \texttt{GetOrbitalCycle}()$\label{alg:get_satellite_period}\;
	$\mathcal{A}_{s}\leftarrow \{k|src_{k}=s, t\leq T_{arv}^{k}\leq t+cyc-1\},\forall s\in SAT$\;
	\textit{/* estimate sunlight duration and task amount. */}\;
	\For{each orbit $i\leftarrow 1,2,\dots,M$}{\label{alg:predict_sunlight_task_begin}
		$sunlit[i]\leftarrow \sum_{s\in ORB_{i}}\sum_{\tau=t}^{t+cyc-1}sun_{s,\tau}$\;
		$task[i]\leftarrow \sum_{s\in ORB_{i}}\sum_{k\in \mathcal{A}_{s}}T_{cp}^{k}$\;
	}\label{alg:predict_sunlight_task_end}
	\For{each orbit $i\leftarrow 1,2,\dots,M$}{
		$w[i]\leftarrow task[i]/\sum_{j=1}^{M} task[j]$\label{alg:task_weight}\;
		$target[i]\leftarrow \texttt{Int}(w[i]*\sum_{j=1}^{M} sunlit[j])-sunlit[i]$\label{alg:setup_target}\;
	}
	\textit{/* assign orbit subset. */}\;
	$idle\leftarrow \{i|task[i]=0\},\forall i,1\leq i\leq M$\;
	\For{each orbit $i\leftarrow 1,2,\dots,M$}{
		\eIf{$target[i]< 0$}{
			$alt\_set[i]\leftarrow \{i\}$\label{alg:self_satisfy}
		}{
			$subset\leftarrow \texttt{Knapsack}(idle, sunlit, target[i])$\label{alg:knapsack}\;
			$alt\_set[i]\leftarrow \{i\}\cup subset$\;
			$idle\leftarrow idle-subset$\;
		}
	}
	\KwRet{$alt\_set$}
\end{algorithm}
\setlength{\floatsep}{0.12cm}

\subsection{Sunlight-Aware Dynamic SEC Task Scheduling Algorithms}
\label{subsec:algorithm}

To solve the SBEO problem, we decompose the problem into three parts based on the \name architecture proposed in \S\ref{subsec:system_overview}: predetermined orbit assignment, on-board offloading selection and processing arrangement. First, we propose a sunlight-aware orbit assignment algorithm (Algorithm~\ref{alg:orbit_assignment}) running in the mission center before task arrival. 
The mission center calculates an alternative orbit subset (denoted as $alt\_set$) for each orbit as the input of the offloading selection algorithm. Satellites can only offload tasks to ground stations or orbits in their alternative subset, which can avoid competition among orbits. Second, we propose an orbit-based offloading algorithm (Algorithm~\ref{alg:offloading_selection}) running in each satellite to decide the processing matrix $\mathcal{X}$ and offloading destination $dst_{k}$, which invokes the processing arrangement algorithm (Algorithm~\ref{alg:processing_arrangement}). Note that the output of SBEO problem $\mathcal{K}_{s}$ can be constructed from $dst_{k}$. Last, the processing arrangement algorithm decides when to process tasks, minimizing the battery energy consumption while avoiding deadline miss. All of the three algorithms can be solved in polynomial time, thus we can solve the SBEO problem in polynomial time via such decomposition.

\noindent
\textbf{Orbit assignment in mission center.} The key ideas 
are summarized as follows: (i) To fairly assign the orbit subsets, the energy capability of each subset should match the task amount. Therefore, we use sunlight duration in a period to represent the energy capability and select the orbit subsets which satisfy the following properties: 
the subsets for orbits generating tasks are disjoint and 
the energy capability proportion of each subset is close to the task proportion. 
(ii) To reduce the complexity of searching suitable subsets, we set up a target capability for each subset according to task amount and transfer the problem to a knapsack problem, which finds an orbit subset that minimizes the gap to target capability. 

Algorithm~\ref{alg:orbit_assignment} shows the details of sunlight-aware orbit assignment algorithm. First, function $\texttt{GetOrbitalCycle}$ calculates the orbital period $cyc$, the time for a satellite to complete one orbit (line~\ref{alg:get_satellite_period}). $\mathcal{A}_{s}$ denotes the tasks generated by satellite $s$ in next period. Then, we predict the sunlight duration and task amount of each orbit in the next period, represented by $sunlit[i]$ and $task[i]$ for orbit $i$ respectively (line~\ref{alg:predict_sunlight_task_begin}-\ref{alg:predict_sunlight_task_end}). $w[i]$ records the task amount proportion of orbit $i$ (line~\ref{alg:task_weight}) and we expect to find a subset with similar energy capability proportion. So we set up the target capability (denoted as $target[i]$) for orbit $i$ based on the task amount proportion. The alternative subset for each orbit must include itself, thus we subtract its energy capability and convert the target capability into an integer for knapsack problem (line~\ref{alg:setup_target}). We use $idle$ to represent the set of orbits without task generation. If the target capability is less than 0, the orbit can self-satisfy the energy requirement, so the alternative set is itself (line~\ref{alg:self_satisfy}). Otherwise, we regard orbits as items associated with weights $sunlit$ and call function $\texttt{Knapsack}$ to select orbits from $idle$ such that the gap to bag capacity $target[i]$ is minimized (line~\ref{alg:knapsack}). In the knapsack problem, the number of items is at most $M$ and bag capacity $target[i]$ is at most $cyc\cdot |SAT|$. Therefore, the time complexity of Algorithm~\ref{alg:orbit_assignment} is $\mathcal{O}(M^2\cdot cyc\cdot |SAT|)$. 

\begin{algorithm}[tb]
    \caption{Orbit-based Offloading} 
    \label{alg:offloading_selection}
	\DontPrintSemicolon
	\KwIn{satellite $src$, task $k$, alternative subset $alt\_set$}
	\KwOut{offloading node $dst_{k}$, processing matrix $\mathcal{X}$}
		\textit{/* offload to ground station. */}\;
		$gs, T_{gs}\leftarrow \texttt{GetAvailableGS}(t)$\label{alg:get_ground_station}\;
		\If{$T_{gs}+z_{k}/Cap(src,gs)\leq T_{ddl}^{k}$}{\label{alg:gs_offloading_begin}
			$T_{gs}\leftarrow T_{gs}+z_{k}/Cap(src,gs)$\label{alg:update_available_time}\;
			$dst_{k}\leftarrow gs$, \textbf{continue}\label{alg:ground_station_offloading}\;
		}\label{alg:gs_offloading_end}
		$\hat{\mathcal{X}}, flag\_sun \leftarrow \texttt{Arrange}(src,t,\mathcal{K}_{src}\cup \{k\})$\;
		\eIf{$flag\_sun==1$}{
			$dst_{k}\leftarrow src$  \textit{/* process locally. */}\label{alg:local_processing}\;
			
		}{
			\textit{/* offload to other satellite. */}\;
			$cnt[]\leftarrow \texttt{GetTaskCounter}()$\label{alg:get_task_counter}\;
			$i\leftarrow \arg\min_{j\in alt\_set[src.orbit]}cnt[j]/sunlit[j]$\label{alg:min_task_sunlight_ratio}\;
			$cnt[i]\leftarrow cnt[i]+T_{cp}^{k}$\label{alg:update_task_counter}\;
			$E[s]\leftarrow \texttt{QueryEnergy}(s),\forall s\in ORB_{i}$\label{alg:query_energy}\;
			$dst_{k}\leftarrow \arg\max_{s\in ORB_{i}}E[s]$\label{alg:select_max_energy}\;
			
		}
		\If{$dst_k==src$}{
			$\mathcal{X}\leftarrow \hat{\mathcal{X}}$ \textit{/* confirm local processing time.*/}\label{alg:local_processing_time}
		}
		\KwRet{$dst_{k}$, $\mathcal{X}$}
	
\end{algorithm}

\noindent
\textbf{Offloading selection in each satellite.} Based on the alternative subset assigned by Algorithm~\ref{alg:orbit_assignment}, the orbit-based offloading algorithm adopts the following ideas: (i) As the connection of GSLs can be predicted, we first check whether tasks can be offloaded to ground stations before deadline to save on-board computation resources. (ii) For satellite offloading, we maintain a task counter for orbits in the alternative subset and try to keep the task amount conforming to their energy capability.

Algorithm~\ref{alg:offloading_selection} shows the details of the orbit-based offloading algorithm. When a new task $k$ arrives at satellite $src$, the offloading manager obtains the nearest available ground station $gs$ and its available time $T_{gs}$ via function $\texttt{GetAvailableGS}$ (line~\ref{alg:get_ground_station}). If the task can be offloaded to $gs$ before deadline, we select $gs$ as the offloading destination and update the available time (line~\ref{alg:gs_offloading_begin}-\ref{alg:gs_offloading_end}). Note that we do not adopt multi-hop ground station offloading due to the re-routing problem caused by GSLs change.
If not, we try to arrange the task locally via Algorithm~\ref{alg:processing_arrangement}, searching for the possible processing matrix $\hat{\mathcal{X}}$ and checking whether the task can be processed completely in sunlight (denoted as $flag\_sun$). If $flag\_sun$ is 1, we process the task locally (line~\ref{alg:local_processing}). Otherwise, we offload the task to other satellite. We obtain the task counter array $cnt$ via function $\texttt{GetTaskCounter}$, which is initialized to 0 in the first timeslot. Then, we select the orbit with minimum ratio of task counter to sunlight duration and update its task counter (line~\ref{alg:min_task_sunlight_ratio}-\ref{alg:update_task_counter}). 
The source satellite sends queries to satellites in the selected orbit and each satellite receiving the query responds with its energy state $E[s]$, which is decided by sunlight duration, rest battery energy and task queue:
\begin{align}
	E[s] = P_{solar}^{s} \sum_{\tau=t}^{t+cyc-1}sun_{s,\tau}+B_{s,t-1}-P_{cp}^{s} \sum_{k\in \mathcal{K}_{s}}T_{cp}^{k}.
\end{align}
Then, the source satellite chooses the one with maximum energy as the offloading node (line~\ref{alg:select_max_energy}). Finally, if the task is processed locally, we adopt the processing arrangement calculated before to avoid recomputation (line~\ref{alg:local_processing_time}).

\noindent
\textbf{Processing arrangement in each satellite.} To arrange the processing time, we apply the deadline first scheme and search for the delay to exploit as less eclipse timeslots. The details of processing arrangement algorithm are shown in Algorithm~\ref{alg:processing_arrangement}. First, we sort the tasks according to their deadlines in ascending order. Then, we calculate the latest time of task $k$ to start processing (denoted as $T_{latest}^{k}$), which guarantees the deadline requirement (line~\ref{alg:cal_latest_time_begin}-\ref{alg:cal_latest_time_end}). Next, we calculate the earliest time of task $k$ to start processing (denoted as $T_{earliest}^{k}$). The indicator $flag\_sun$ is initialized to 1 and the earliest time of the first task is initialized to current time $t$. Then, we search for the next sunlit time $T_{sunlit}^{k}$ (line~\ref{alg:next_sunlight_phase}). If the next sunlit time is before the latest task processing time, we arrange the task to be processed from $T_{sunlit}^{k}$ to save battery energy (line~\ref{alg:arrange_at_sunlight}). Otherwise, we arrange the task to be processed from $T_{latest}^{k}$ to guarantee the deadline requirement (line~\ref{alg:arrange_at_latest}) and set $flag\_sun$ to 0 (line~\ref{alg:update_cost}). At the end of each loop, we update 
the earliest processing time of next task (line~\ref{alg:update_earliest_time}).

\begin{algorithm}[tb]
    \caption{Processing Arrangement} 
    \label{alg:processing_arrangement}
	\SetKwFunction{FArrange}{Arrange}
	\SetKwProg{Fn}{Function}{:}{\KwRet}
	\Fn{\FArrange{$s$,$t$,$\mathcal{K}_{s}$}}{
    	$\texttt{Sort}(\mathcal{K}_{s})$ based on $T_{ddl}^{k}$ in ascending order
		$T_{latest}^{|\mathcal{K}_{s}|}\leftarrow T_{ddl}^{|\mathcal{K}_{s}|}-T_{cp}^{|\mathcal{K}_{s}|}$\label{alg:cal_latest_time_begin}
		\For{$k\leftarrow |\mathcal{K}_{s}|-1,|\mathcal{K}_{s}|-2,\dots,1$}{
			{$T_{latest}^{k}\leftarrow \min(T_{ddl}^{k},T_{latest}^{k+1})-T_{cp}^{k}$}
		}\label{alg:cal_latest_time_end}
		$flag\_sun\leftarrow 1,\ T_{earliest}^{1}\leftarrow t$
		\For{$k\leftarrow 1,2,\dots,|\mathcal{K}_{s}|$}{
			$T_{sunlit}^{k}\leftarrow \min\{\tau|sun_{s,\tau}=1,\tau\geq T_{earliest}^{k}\}$\label{alg:next_sunlight_phase}
			\eIf{$T_{sunlit}^{k}\leq T_{latest}^{k}$}{
				$x_{k,\tau}\leftarrow 1,\forall \tau, T_{sunlit}^{k}\leq \tau< T_{sunlit}^{k}+T_{cp}^{k}$\label{alg:arrange_at_sunlight}
			}{
				$x_{k,\tau}\leftarrow 1,\forall \tau, T_{latest}^{k}\leq \tau< T_{latest}^{k}+T_{cp}^{k}$\label{alg:arrange_at_latest}
				$flag\_sun\leftarrow 0$\label{alg:update_cost}
			}
			$T_{earliest}^{k+1}\leftarrow \min(T_{sunlit}^{k},T_{latest}^{k})+T_{cp}^{k}$\label{alg:update_earliest_time}
		}
		\KwRet{$\mathcal{X}, flag\_sun$}
	}
\end{algorithm}

\section{Performance Evaluation}
\label{sec:evaluation}
In this section, we implement an SEC testbed and evaluate the performance of \name to illustrate its effectiveness. First, we compare the DoD of satellite batteries and task completion time with the other state-of-the-art strategies. Second, we explore the performance under four seasons as the revolution of the earth can affect the sunlit ratio defined in \S\ref{subsec:observation_sunlit_sufficient_space_edge}. Third, we apply the strategies to different constellations to explore the impact of constellation parameters. Finally, we setup various workloads to illustrate the robustness of \name via tuning the processing capability and task type. 

\subsection{Environment Setup}


\begin{figure}[h]
	\centering
	\includegraphics[width=0.99\linewidth]{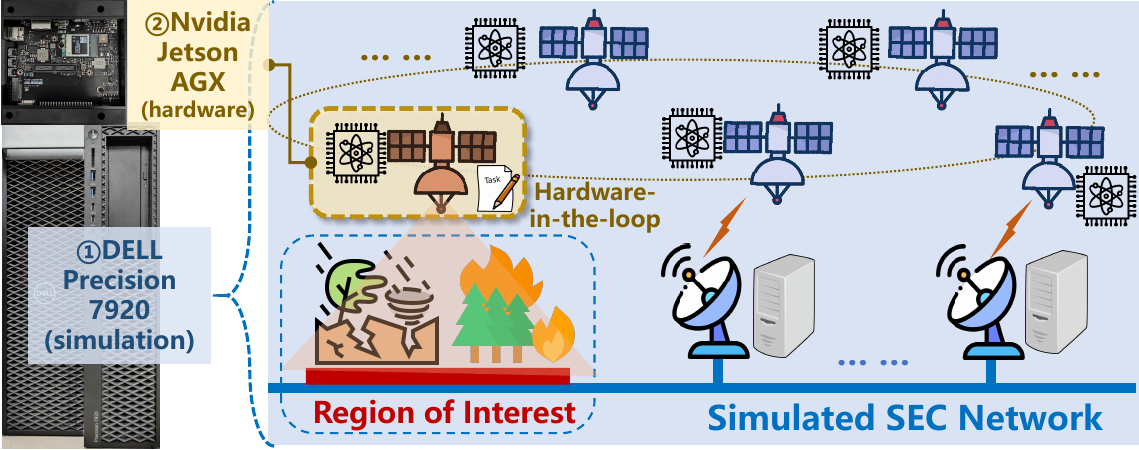}
	\vspace{-0.2in}
	\caption{Our hardware-in-the-loop testbed combines: (1) a Nvidia Jetson low-power edge computing hardware to mimic a real satellite edge, and (2) a high-end server simulating a large-scale SEC network, distributed ground stations \etc}
	\label{fig:prototype_architecture}
\end{figure} 

\noindent
\textbf{Prototype implementation.} We build a data-driven hardware-in-the-loop SEC testbed based on StarryNet~\cite{lai2023starrynet}, a recent container-based satellite network emulator. The SEC environment is deployed on a Dell Precision 7920 Tower Workstation connected with a Jetson AGX Orin Developer Kit as shown in Fig.~\ref{fig:prototype_architecture}. 
The Developer Kit works as an SEC node, running machine learning models to evaluate the task completion time and energy consumption.
Based on the +Grid~\cite{bhattacherjee2019network} structure and the trajectory of the satellite simulated by the Developer Kit, it establishes virtual links with its adjacent satellites and ground facilities dynamically.

\begin{figure*}[tp]
	\centering
	\subfloat[DoD distribution of satellites.]{
		\includegraphics[width=0.32\linewidth]{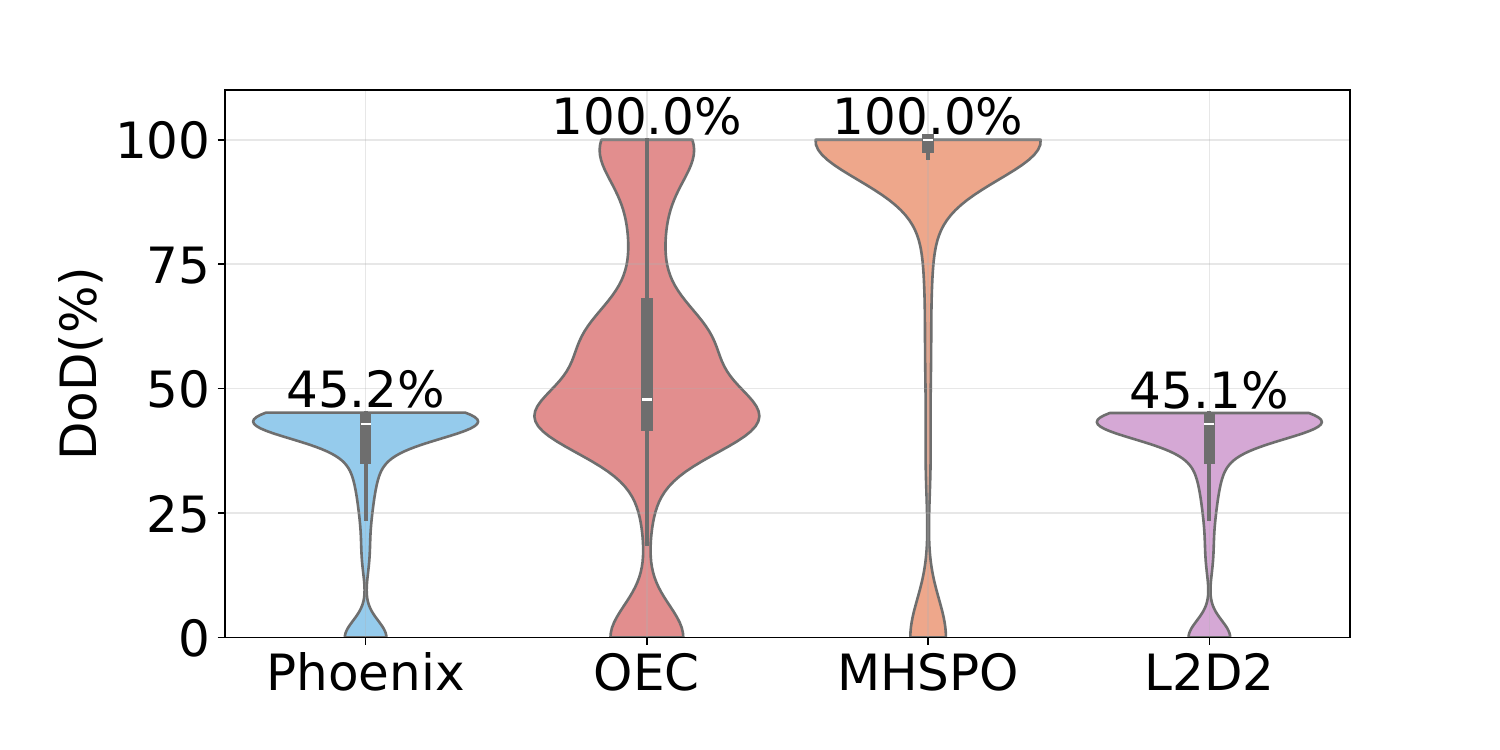}
		\label{fig:starlink_simulation_dod}
	}
	\subfloat[The percentage of scheduling decision.]{
		\includegraphics[width=0.33\linewidth]{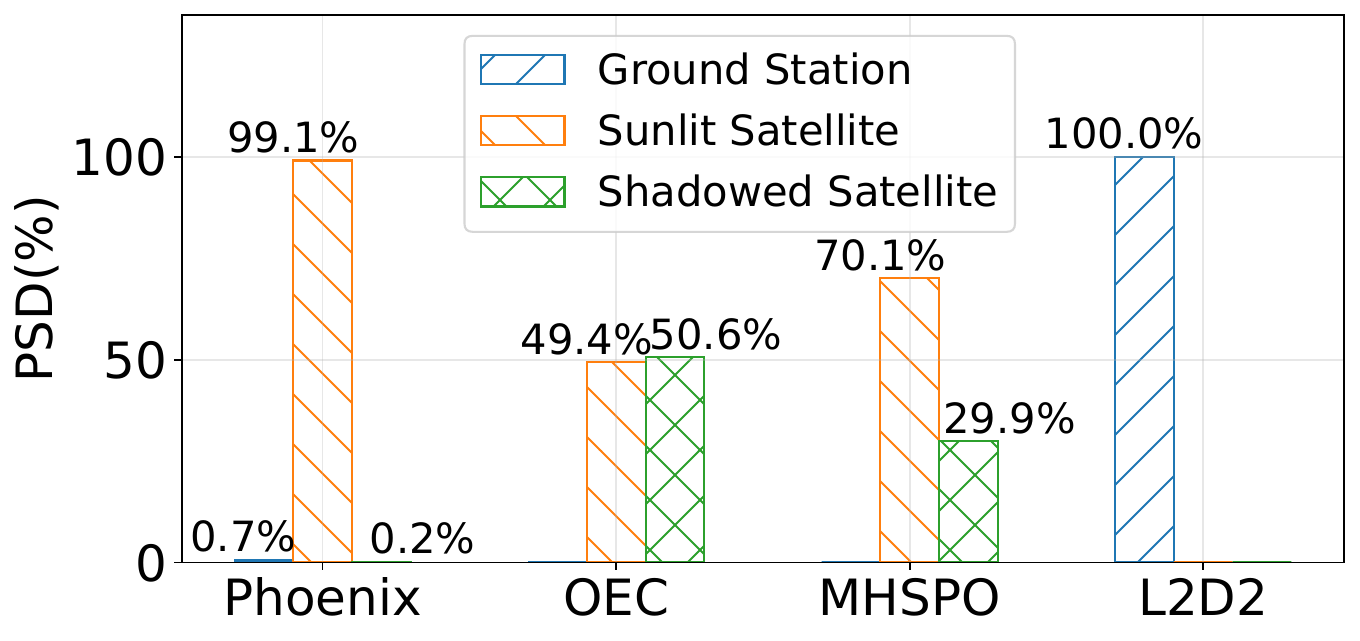}
		\label{fig:starlink_sunlit_ratio}
	}
	\subfloat[CDF of task completion time with log scale.]{
		\includegraphics[width=0.3\linewidth]{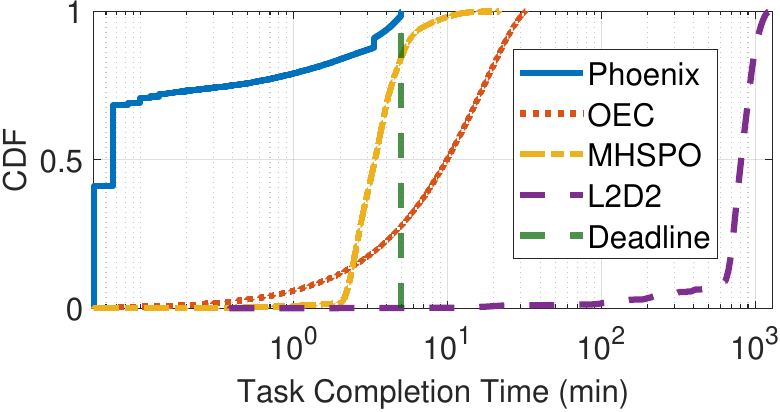}
		\label{fig:starlink_task_completion_time}
	}
	\vspace{-0.05in}
	\caption{Performance comparison of different offloading strategies.}
	\vspace{-0.25in}
\end{figure*}

\noindent
\textbf{LEO constellation settings.} We conduct extensive simulation driven by real-world information, including two kinds of LEO constellations: inclined orbit constellation (Starlink~\cite{fcc_starlink}) and polar constellation (OneWeb~\cite{fcc_oneweb}). 
Following \cite{vasisht2021l2d2}, we use the ground station locations collected by SatNOGS~\cite{satnogs}. 
For computation, we adjust the power level of Jetson AGX Orin Developer Kit to 30W/50W/60W, providing different computation capabilities. 
Based on the existing hardwares~\cite{radiosat,cubelct}, we set the power of GSL/ISL to 16W/10W respectively. Following~\cite{denby2020orbital}, we set the basic power to 4W. 
And we set the power generated by solar panels to 120W and the battery volume to 60Wh, which can offer sufficient energy. 
For transmission, we set the capacity of GSL/ISL to 100Mbps/1Gbps respectively.

\noindent
\textbf{SEC tasks and datasets.} We select ship detection~\cite{yang2022algorithm,wei2020hrsid} and wildfire segmentation~\cite{rashkovetsky2021wildfire} as the SEC tasks. For ship detection, we apply YOLO~\cite{redmon2016you} to dataset~\cite{ship_detection_dataset} and select Atlantic Ocean as the RoI. Satellites continuously capture images every second, with the resolution of $10K\times 10K$ pixels~\cite{denby2023kodan}. Each image is expected to be processed within 5 minutes~\cite{kerr2021eo}.
Based on our measurement, the processing time of an image for ship detection is 10s/5s/3s under 30W/50W/60W respectively.  
For wildfire segmentation, we apply U-Net~\cite{ronneberger2015u} to dataset~\cite{t9gn-y009-20} and select Amazon Rainforest as the RoI. The imaging interval is set to 5 seconds and the processing time of an image is 120s/67s/51s under 30W/50W/60W respectively. Other parameters are the same as ship detection task.

\noindent
\textbf{Comparison objects and metrics.} We implement three state-of-the-art offloading schemes for comparison: (i) OEC~\cite{denby2020orbital}, which processes the tasks with an intra-orbit pipeline; (ii) MHSPO~\cite{zhang2023energy}, an energy-efficient satellite peer offloading scheme, and (iii) L2D2~\cite{vasisht2021l2d2}, which offloads tasks to geo-distributed ground stations. We regard L2D2 as the baseline because it doesn't consume any \emph{computation energy}, and thus has the lowest energy consumption. We use DoD, battery lifetime and task finish time as metrics to present the effectiveness of \name.

\subsection{DoD and Task Completion Time Comparison}

We first compare the performance under the configuration of Starlink constellation, using 60W power level for ship detection. As shown in Fig.~\ref{fig:starlink_simulation_dod}, \name is close to L2D2 and reduces the maximum DoD by 54.8\% as compared with OEC and MHSPO. Note that there are some satellites with 0\% DoD because these satellites can keep illuminated by sun without consuming the battery energy. As tasks can be processed in ground stations, sunlit satellites or shadowed satellites, Fig.~\ref{fig:starlink_sunlit_ratio} plots the percentage of scheduling decisions (PSD), which indicates the ratio of three types of nodes selected to process tasks. We can see that \name outperforms other on-board computation strategies from two aspects: (i) \name can cooperatively exploit the communication capability of ground stations and on-board computation capability of satellites; (ii) \name can exploit the sunlit satellites to process tasks (99.1\% tasks are processed in sunlit satellites) while reducing the consumption of shadowed satellites to save the energy of battery (only 0.2\% tasks are processed in shadowed satellites). 
As shown in Fig.~\ref{fig:starlink_task_completion_time}, \name can satisfy the deadline requirement while OEC, MHSPO and L2D2 may miss deadline.  
As L2D2 offloads tasks to ground stations directly without on-board computing, it consumes the least battery energy, but the task completion time is very high due to the downlink bottleneck. Overall, \name can not only reduce the DoD of batteries, achieving the near-optimal performance, but also satisfy the deadline requirement.

\subsection{Impact of Seasons and Battery Lifetime Extension}
\begin{figure}[tp]
	\centering
	\subfloat[Average DoD under four seasons.]{
		\includegraphics[width=0.48\linewidth]{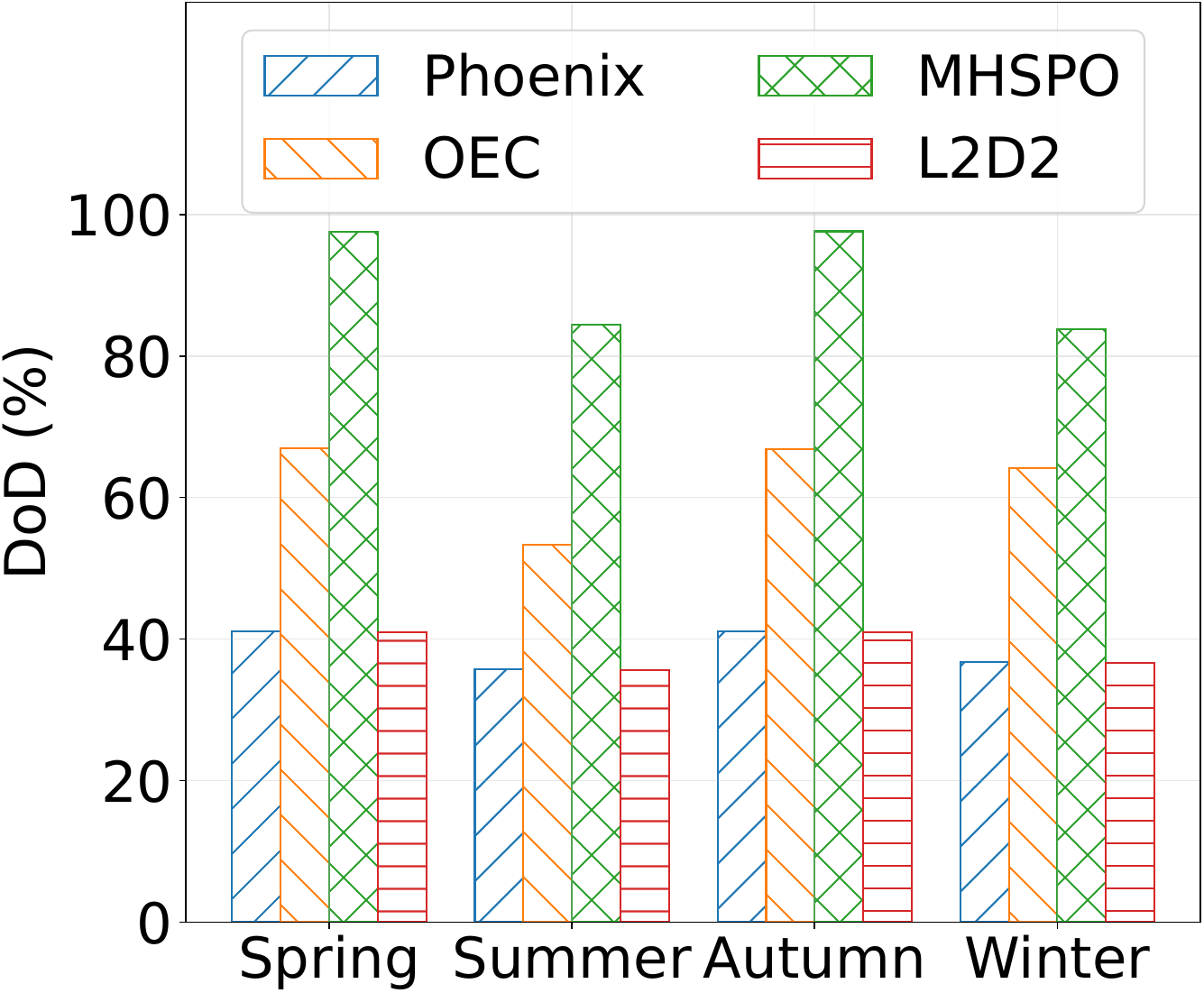}
		\label{fig:dod_different_seasons}
	}
    \subfloat[Estimated battery lifetime.]{
		\includegraphics[width=0.45\linewidth]{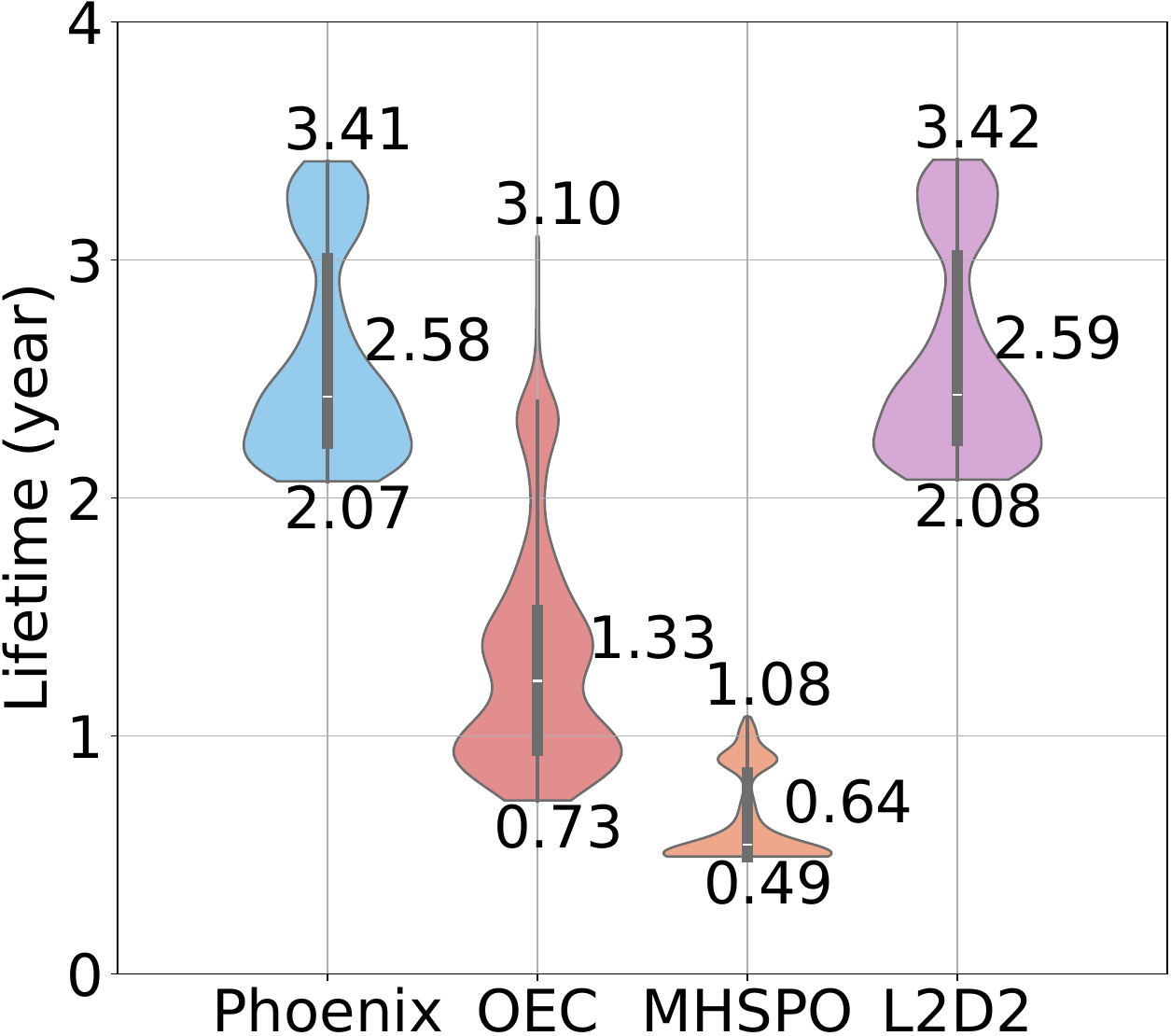}
		\label{fig:starlink_simulation_battery_lifetime}
	}
    \vspace{-0.05in}
    \caption{Performance under Starlink configuration in 2023.}
    \vspace{-0.15in}
\end{figure}    
\begin{figure}[tp]
    \centering
    \subfloat[Average DoD under four seasons.]{
		\includegraphics[width=0.48\linewidth]{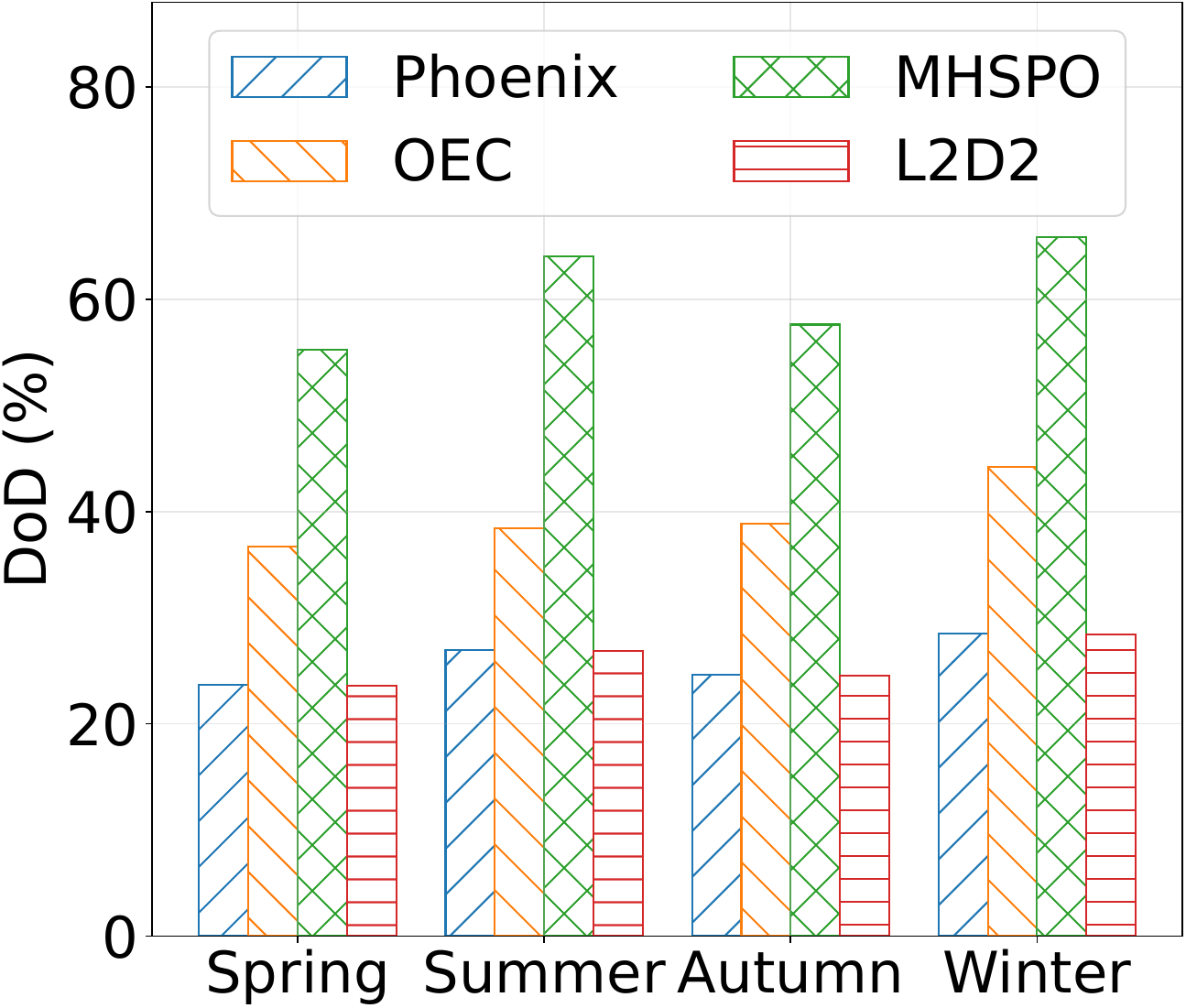}
		\label{fig:oneweb_dod_seasons}
	}
	\subfloat[Estimated battery lifetime.]{
		\includegraphics[width=0.47\linewidth]{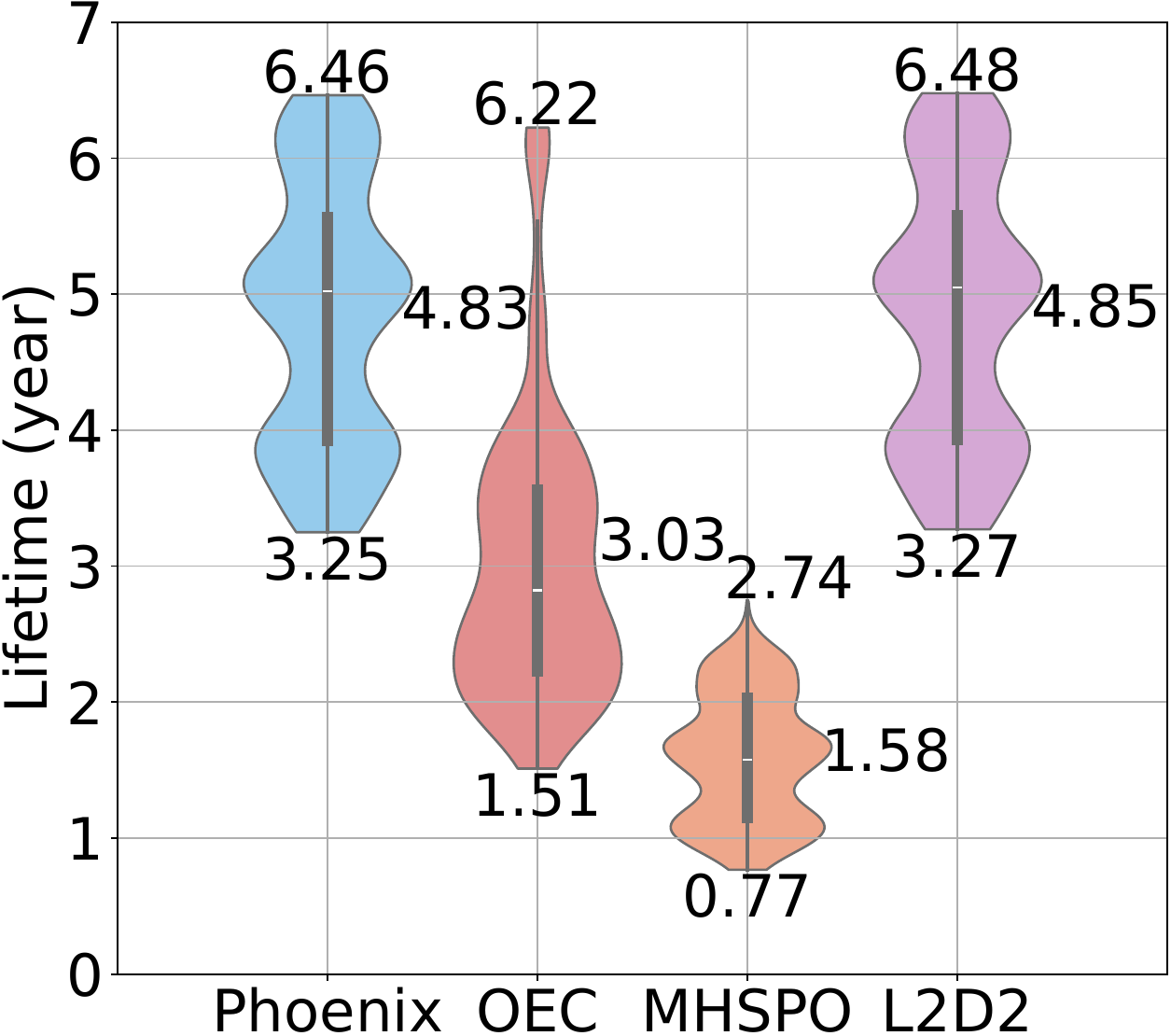}
		\label{fig:oneweb_battery_lifetime}
	}
    \caption{Performance under OneWeb configuration in 2023.}
    \vspace{-0.05in}
\end{figure}

As the earth revolves around the sun, the sunlit ratio of each satellite may change over a year. Thus, we repeat the above experiments under the configuration of four seasons, comparing the average DoD in Fig.~\ref{fig:dod_different_seasons} and estimated lifetime in Fig.~\ref{fig:starlink_simulation_battery_lifetime}. As shown in Fig.~\ref{fig:dod_different_seasons}, \name is close to L2D2 and outperforms OEC/MHSPO in any season. Meanwhile, we find that the DoD of all strategies in summer and winter is lower than that in spring and autumn.
This is because the angles between some inclined orbits and sunlight are nearly vertical in summer and winter, which can offer larger sunlit ratio. 
This phenomenon also indicates that the sunlight is an important factor in energy optimization. Following the life model in~\cite{mallon2017analysis}, we estimate the battery lifetime of each satellite when applying different strategies. As shown in Fig.~\ref{fig:starlink_simulation_battery_lifetime}, \name is close to L2D2 and can prolong the satellite lifespan up to $2.9\times/5.3\times$ as compared with OEC/MHSPO. This is because \name is sunlight-aware and exploits the sunlit satellites to process tasks, which can significantly reduce the battery energy consumption.

\subsection{Impact of Constellation Parameters}

The sunlit ratio is also correlated to the parameters of constellations (\eg inclination, altitude) as mentioned in \S\ref{subsec:observation_sunlit_sufficient_space_edge}. We simulate OneWeb constellation in our testbed and compare the performance of four strategies. As shown in Fig.~\ref{fig:oneweb_dod_seasons}, \name is close to L2D2 and outperforms OEC/MHSPO by up to 15.7\%/37.4\% on average DoD. 
Different from inclined orbit, the DoD in summer and winter is larger than that in spring and autumn. This is because polar orbit gets more sunlit ratio in spring and autumn, which is opposite to the inclined orbit.
Fig.~\ref{fig:oneweb_battery_lifetime} plots the lifetime of satellites, which shows that \name can achieve longer lifetime to $2.3\times/4.4\times$ as compared with OEC/MHSPO. When applying \name under different constellation configurations, OneWeb can achieve better performance than Starlink with lower DoD (12.8\% on average) and longer lifetime ($1.9\times$ on average) for two key reasons: (i) polar orbit can experience longer sunlight duration than inclined orbit; (ii) the altitude of OneWeb is higher than that of Starlink, which can obtain higher sunlit ratio.

\subsection{Impact of Various Capabilities and Workloads}

Finally, we adjust the power of Developer Kit from 30W to 60W and apply wildfire segmentation task to explore the performance under various computation capabilities and workloads. TABLE~\ref{tbl:average_dod_diff_power} shows the average DoD of each strategy under different power levels and workloads. Results show that \name is close to L2D2 and outperforms others under various computation capabilities and workloads, which indicates the robustness of \name. The DoD of L2D2 remains unchanged due to no on-board processing. When the power increases, the DoD becomes smaller for \name, which is somewhat counterintuitive. This is because the processing time of tasks gets shorter with strengthened computation capability. The energy is relative to both power and time, thus the energy consumed by a ship detection (wildfire segmentation) task is 300J/250J/180J (3600J/3350J/3060J) under 30W/50W/60W respectively. This inspires us that even if we promote the computation capability, the battery energy consumption can be reduced via proper task scheduling.

\begin{table}[t]
	\begin{center}
		\caption{Average DoD under different processing capabilities and two types of workloads.}
		\label{tbl:average_dod_diff_power}
		\begin{tabular}{|c|c|c|c|c|c|} 
			\hline
			\multicolumn{2}{|c|}{\diagbox{\textbf{Power}}{\textbf{DoD(\%)}}{\textbf{Strategy}}} & \textbf{\name}& \textbf{OEC} & \textbf{MHSPO} & \textbf{L2D2}\\
			\hline
			\multirow{3}{*}{\makecell[c]{Ship\\Detection}} & 30W & 43.1 & 47.5 & 59.3 & 35.7\\
			& 50W & 36.0 & 53.2 & 80.5 & 35.7 \\
			& 60W & 35.7 & 53.3 & 84.4 & 35.7\\
			\hline
			\multirow{3}{*}{\makecell[c]{Wildfire\\Segmentation}} & 30W & 42.1 & 42.7 & 58.2 & 35.7\\
			& 50W & 41.3 & 47.0 & 79.0 & 35.7 \\
			& 60W & 38.4 & 48.7 & 83.9 & 35.7\\
			\hline
		\end{tabular}
	\end{center}
	\vspace{-0.1in}
\end{table}
\section{Related Work}
\label{sec:related_work}


\noindent
\textbf{Efficient network delivery for big in-orbit data.} Emerging satellites with evolved remote sensing capabilities are widely used in many applications such as the earth surveillance and disaster monitoring. 
A number of recent efforts have studied the approaches for accelerating space data delivery and optimizing the task completion time~\cite{vasisht2021l2d2,lai2021orbitcast,lyu2023falcon}. L2D2~\cite{vasisht2021l2d2} is a space data download scheme which uses commodity hardware to offer low latency and robust download. OrbitCast~\cite{lai2021orbitcast} is a hybrid space data delivery architecture that collaboratively leverages LEO satellites and geo-distributed ground stations to fast forward space data. However, with the increasing resolution of emerging on-board sensors, the amount of space data has also increased exponentially in recent years. Downloading all space data to the ground requires massive amounts of bandwidth and storage in satellite, which induces large downloading latency. 

\noindent
\textbf{SEC for in-orbit data processing.} Other efforts investigated the feasibility of leveraging edge-like computation capabilities on emerging satellites to directly process data in orbit~\cite{denby2020orbital,denby2023kodan}. This technique, known as space/orbit edge computing, can identify and discard unnecessary information among the big space data. Network efficiency is improved since only high-value data will be downloaded to the ground. However, these works ignore the energy challenge caused by the additional workload of in-orbit processing. More recently, MHSPO~\cite{zhang2023energy} leverages Lyapunov optimization to optimize energy consumption for SEC networks by offloading tasks to peer satellites. The fundamental difference between MHSPO and \name is that MHSPO ignores the time-varying sunlight states of LEO satellites, which weakens the effectiveness of battery energy optimization for SEC networks.


\noindent
\textbf{Task scheduling in mobile edge computing.} Energy-efficient task scheduling~\cite{guo2016energy,kwak2015dream,chen2015efficient,deng2014computation} has been well studied in terrestrial networks, which exploit dynamic CPU frequency scaling, network interface selection and transmission power allocation techniques to adaptively make offloading decisions and control energy consumption. However, the large amount of tasks makes them difficult to scale the satellite computation and communication capability.
Recent learning-based works~\cite{ren2023feat,tang2020deep} have applied deep reinforcement learning in mobile devices to select offloading actions. However, the action space explodes among large-scale satellite agents and selecting actions via deep neural network consumes extra energy.

\section{Conclusion}
\label{sec:conclusion}

In recent years, \emph{space edge computing~(SEC)} is becoming a new computation paradigm for future integrated space and terrestrial networks.
In this paper, we present \name, an energy-efficient task scheduling framework for futuristic SEC networks. \name exploits a number of \emph{sunlit edges} for on-board task processing. In particular, we propose a series of sunlight-aware scheduling algorithms to reduce battery energy consumption. We implement a \name prototype and conduct experiments on the SEC environment. Evaluations demonstrate that as compared to the state-of-the-art solutions, \name can reduce the DoD by up to 54.8\% and prolong the battery lifetime to $2.9\times$ while guaranteeing task deadline.
\section*{Acknowledgment}
This work was supported by the National Key R\&D Program of China (No. 2022YFB3105202) and National Natural Science Foundation of China~(NSFC No. 62132004 and No. 62372259). Qian Wu is the corresponding author.

\bibliographystyle{IEEEtran}
\bibliography{reference}

\end{document}